# Anisotropic upper critical field of BiS$_2$-based superconductor LaO$_{0.5}$F$_{0.5}$BiS$_2$

Yoshikazu Mizuguchi[1], Atsushi Miyake[2], Kazuto Akiba[2], Masashi Tokunaga[2], Joe Kajitani[1], Osuke Miura[1]

1) Department of Electrical and Electronic Engineering, Tokyo Metropolitan University, 1-1, Minami-osawa, Hachioji, Tokyo 192-0397, Japan
2) The Institute for Solid-State Physics (ISSP), the University of Tokyo, 5-1-5, Kashiwanoha, Kashiwa, Chiba 277-8581 Japan

Abstract

Superconducting properties under high magnetic fields for the novel BiS$_2$-based layered superconductor LaO$_{0.5}$F$_{0.5}$BiS$_2$ (polycrystalline sample prepared using the high-pressure synthesis method) have been investigated. The anisotropy of the upper critical field was discussed by analyzing the temperature dependence of the temperature derivative of resistivity. The temperature dependence of upper critical field showed an anomalous behavior with a characteristic magnetic field of 8 T. We revealed that the anomalous behavior was caused by the existence of three kinds of upper critical field, $\mu_0 H_{C2}^{max}$, $\mu_0 H_{C2}^{mid}$ and $\mu_0 H_{C2}^{min}$ in LaO$_{0.5}$F$_{0.5}$BiS$_2$. The lowest upper critical field $\mu_0 H_{C2}^{min}$ is regarded as the $\mu_0 H_{c2}$ where the superconducting states of the grains with an orientation of $H // c$ are suppressed. The $\mu_0 H_{C2}^{max}$ and $\mu_0 H_{C2}^{mid}$ are regarded as the $\mu_0 H_{c2}$ where the superconducting states of the grains with an orientation of $H // ab$ are suppressed. The difference between the $\mu_0 H_{C2}^{max}$ and $\mu_0 H_{C2}^{mid}$ could be explained by the anisotropy of superconducting states within the Bi-S plane. The estimated anisotropy parameter of the upper critical fields was about 7.4 for the high-pressure-synthesized LaO$_{0.5}$F$_{0.5}$BiS$_2$ polycrystalline sample. Since the value was clearly lower than the anisotropy parameter of over 30 observed in the previous report on the LaO$_{0.5}$F$_{0.5}$BiS$_2$ single crystals grown under ambient pressure, we concluded that a larger anisotropy of superconductivity was not essential for the enhancement of $T_c$, and lower anisotropy might be important for a higher $T_c$ in LaO$_{1-x}$F$_x$BiS$_2$.

## 1. Introduction

Layered superconductors have drawn many research attentions because of the appearance of unconventional superconductivity and a high transition temperature ($T_c$). So far, two kinds of high-$T_c$ layered superconductors, cuprates and Fe-based superconductors, have been discovered [1,2]. Both the high-$T_c$ superconductors possess a crystal structure composed of an alternate stacking of common superconducting layers and spacer layers. Recently, we have discovered novel layered superconductors with the $BiS_2$ superconducting layers [3,4]. This family possesses a crystal structure composed of an alternate stacking of the common $BiS_2$ superconducting layers and the various spacer layers such as $RE_2O_2$ (RE = La, Ce, Pr, Nd and Yb) [4-11], $Sr_2F_2$ [12,13] and $Bi_4O_4(SO_4)_{1-x}$ layers [3]. Their crystal structure at room temperature and under ambient pressure is categorized into the tetragonal space group of *P4/nmm*. Basically, a parent phase of the $BiS_2$ family is a semiconductor with a band gap. The $BiS_2$-based materials become metallic with electron doping into the $BiS_2$ layers and then exhibit superconductivity at low temperatures. In the superconducting phases, two Bi-6$p$ orbitals of $6p_x$ and $6p_y$ are essential for the evolution of superconducting states [14-20].

So far, the highest $T_c$ in the $BiS_2$-based family is $T_c^{onset}$ ~ 11 K of $LaO_{0.5}F_{0.5}BiS_2$. The superconducting properties of $LaO_{0.5}F_{0.5}BiS_2$ are highly sensitive to application of pressure. Although as-grown $LaO_{0.5}F_{0.5}BiS_2$ prepared by solid-state reaction shows a superconducting transition with $T_c$ ~ 3 K [2], the $T_c$ can be enhanced to over 10 K by applying high pressure [21-23]. The higher-$T_c$ phase of $LaO_{0.5}F_{0.5}BiS_2$ can be obtained by annealing the as-grown $LaO_{0.5}F_{0.5}BiS_2$ sample under high pressure of 2 GPa. Namely, the higher-$T_c$ (high-pressure) structure could be memorized even after the sample is put at ambient pressure after the high-pressure annealing [2,23]. One-step high-pressure synthesis can also provide a higher-$T_c$ phase of $LaO_{0.5}F_{0.5}BiS_2$ under the optimal synthesis condition of 2 GPa and 700 °C [24]. Crystal structure analysis revealed that the higher-$T_c$ phase of $LaO_{0.5}F_{0.5}BiS_2$ can be induced by uniaxial lattice contraction along the *c* axis [25]. Furthermore, Tomita *et al.* reported that the lattice contraction under high pressure could be explained by a structural transition from tetragonal to monoclinic on the basis of analysis of the x-ray diffraction patterns under high pressure and optimization of the atomic positions by using density functional theory calculations [23]. In this article, we show a largely anisotropic magnetic field-temperature (*H-T*) phase diagram of superconductivity in the higher-$T_c$ phase of $LaO_{0.5}F_{0.5}BiS_2$.

## 2. Experimental details

A polycrystalline sample of $LaO_{0.5}F_{0.5}BiS_2$ was prepared using high-pressure synthesis with a cubic-anvil-type 180 ton press. The starting materials are the $La_2S_3$ (99.9 %), $Bi_2O_3$ (99.99 %),

BiF$_3$ (99.9 %) and Bi$_2$S$_3$ powders and the Bi (99.99 %) grains. The Bi$_2$S$_3$ powder was prepared by reacting the Bi and S (99.99 %) grains. The mixture with a nominal composition of LaO$_{0.5}$F$_{0.5}$BiS$_2$ was well-mixed and pressed into a pellet. The sample was sealed into a BN crucible and inserted into a high-pressure cell composed of a carbon heater, pyrophyllite cube and electrodes. The high-pressure cell was pressed with a pressure of 2 GPa and heated at 700 ºC for 1 hour. The temperature dependence of the electrical resistivity ($\rho$-$T$) was measured using a four-terminal method under magnetic fields up to 14 T. The magnetic field dependence of the electrical resistivity ($\rho$-$H$) in pulsed magnetic fields of up to ~ 56 T was measured by utilizing a non-destructive pulsed magnet using the four-terminal method. The duration of the pulsed magnetic field was about 36 msec. In the electrical resistivity measurements, the direction of current was parallel to the applied magnetic fields. The temperature dependence of the magnetic susceptibility after both zero-field cooling (ZFC) and field cooling (FC) was measured using a superconducting interference devise (SQUID) magnetometer with applied magnetic field up to 3 T.

3. Results and discussion

Figure 1(a) shows the temperature dependence of the magnetic susceptibility for LaO$_{0.5}$F$_{0.5}$BiS$_2$ under a field of 0.5 mT. Below 11 K, diamagnetic signals corresponding to the evolution of superconducting states are observed. The irreversible temperature ($T_c^{irr}$), which is a bifurcation point between the ZFC and FC curves, is 8.5 K. Figure 1(b) displays the temperature dependence of the FC susceptibility for LaO$_{0.5}$F$_{0.5}$BiS$_2$ with applied fields of 0.5 mT, 0.5 T and 1 T. It is found that the superconducting states can be destroyed by applying a relatively small field of 1 T. At the same time, we clearly note that the onset of $T_c$ is robust against the application of high magnetic fields.

Figure 2(a) shows the temperature dependence of electrical resistivity for LaO$_{0.5}$F$_{0.5}$BiS$_2$ under magnetic fields up to 14 T. The $T_c^{onset}$ and $T_c^{zero}$ under 0 T are 10.8 K and 7.7 K, respectively. Here, we define the $T_c^{onset}$ as a temperature where resistivity begins to decrease in the $\rho$-$T$ measurements. Figure 2(b) shows the enlargement of the temperature dependence of the resistivity near the $T_c^{onset}$. With increasing magnetic field, the $T_c^{zero}$ is largely suppressed, and the zero-resistivity state is not observed above 3 T within a temperature range of $T > 2$ K. In contrast, the kink observed at $T_c^{onset}$ under 0 T is robust against the application of high magnetic fields and still observable at 14 T. As will be discussed later, the $\rho(T)$ curves above 6 T show additional anomaly (see Fig. 2(b)). To obtain further information under higher magnetic fields, we performed the $\rho$-$H$ measurements in pulsed magnetic fields up to 56 T. Figure 3 shows the field dependence of the electrical resistivity at 1.4 K, 2.4 K, 3.4 K and 4.2 K. The onset point ($\mu_0 H_{c2}$ (pulse)) of the superconducting transition is estimated using a criterion of 99 % of the normal state resistivity. Above $\mu_0H_{c2}$, the resistivity is almost constant as a function of fields. The $\mu_0H_{c2}$ is estimated to be 18.7 T, 15.4 T, 12.7 T and 10.7 T for 1.4 K, 2.4 K, 3.4 K and 4.2 K, respectively.

To analyze the obtained data of the resistivity under various magnetic fields, we plotted the $\mu_0H_{c2}$ and $\mu_0H_{irr}$ estimated using the $T_c^{onset}$ and $T_c^{zero}$ in the $\rho$-$T$ measurements in Fig. 4. The $\mu_0H_{c2}$ estimated from the $\rho$-$H$ measurements with pulsed magnetic fields is also plotted in Fig. 4. The $\mu_0H_{c2}$ (pulse) agrees with the $H_{c2}$ ($\rho$-$T$). It is found that the $\mu_0H_{irr}$ is much lower than the $\mu_0H_{c2}$. This can be understood by the large anisotropy of superconductivity in the BiS$_2$-based superconductors. Recently, Nagao *et al.* reported that the anisotropy of superconductivity in REO$_{1-x}$F$_x$BiS$_2$ was quite large and was relevant to those of cuprate superconductors [27,28]. The sample used in this study is polycrystalline; hence, the grain orientation should be random. Generally in the layered superconductors with large anisotropy of superconductivity, the superconducting states in the grains whose $c$ axis is parallel to the applied magnetic fields are strongly suppressed with applying high magnetic fields. It is conceivable that the large suppression of the fraction of the FC susceptibility at high magnetic fields is caused by the large anisotropy of superconductivity in LaO$_{0.5}$F$_{0.5}$BiS$_2$. In other words, the superconducting states for some grains are suppressed by applying magnetic fields.

As shown in Fig. 4, the temperature dependence of the $\mu_0H_{c2}$ shows an anomalous curve with a characteristic field of ~ 8 T. To clarify the anomalous behavior of the $\mu_0H_{c2}$, we analyzed the anisotropy of the $\mu_0H_{c2}$ using the method proposed by Bud'ko *et al.* [29]. They showed that the anisotropy of the $\mu_0H_{c2}$ for the randomly oriented powder samples of layered superconductors can be estimated by analyzing the temperature dependence of the temperature derivative of magnetization ($dM/dT$). In the method, the maximum upper critical field $\mu_0H_{c2}^{max}$, which could be regarded as the $H_{c2}$ for the grains with a grain orientation of $H//ab$, can be estimated from the onset of $dM/dT$; we call the onset temperature $T^{max}$. The minimum upper critical field $H_{c2}^{min}$, which can be regarded as the $\mu_0H_{c2}$ for the grains with $H//c$ can be estimated from the kink in the $dM/dT$-$T$ curve; we call this temperature $T^{min}$.

The temperature dependence of the $dM/dT$ for LaO$_{0.5}$F$_{0.5}$BiS$_2$ under various magnetic fields of 0.5 T, 1 T, 2 T and 3 T is shown in Fig. 5(a-d), respectively. As represented in Fig. 5(b), we clearly observe the $T^{max}$ and $T^{min}$ as observed in the MgB$_2$ and LuNi$_2$B$_2$C superconductors [29]. However, the signal to noise ratio (S/N) becomes smaller at high magnetic fields due to the large suppression of bulk characteristics of superconductivity. Therefore, we discuss the $\mu_0H_{c2}^{max}$ and $\mu_0H_{c2}^{min}$ by analyzing the temperature derivative of the electrical resistivity to clarify the anomalous behavior of the $\mu_0H_{c2}$-$T$ curve under high magnetic fields. As shown in Fig. 5(a-d), it was confirmed that the $T^{max}$ and $T^{min}$ estimated using the $dM/dT$-$T$ curve almost correspond to those estimated using the $d\rho/dT$-$T$ curve under all magnetic fields up to 3 T.

Furthermore, the S/N ratio of the $d\rho/dT$-$T$ curve is clearly better than that of the $dM/dT$-$T$ curve. Therefore, we here investigate the anisotropy of the $\mu_0H_{c2}$ by analyzing the $d\rho/dT$-$T$ data.

As shown in Fig. 5(a-d), both the $T^{max}$ and $T^{min}$ shift to lower temperatures with increasing magnetic field. In particular, the $T^{min}$ is rapidly suppressed with increasing field and becomes below 2 K above 4 T. Therefore, the $\mu_0H_{c2}^{min}(0\ K)$ is expected to be less than 5 T. At above 5 T, we observed an additional inflection point in the $d\rho/dT$-$T$ curves. Figure 6(a) shows the $d\rho/dT$-$T$ and $\rho$-$T$ curves for LaO$_{0.5}$F$_{0.5}$BiS$_2$ under a magnetic field of 8 T. The onset point, $T^{max}$, is still above 9 K. The $T_c^{onset}$ ($\rho$-$T$) is corresponding to the temperature where the value of the $d\rho/dT$ equals to zero. The inflection point appears at ~5.3 K; here, we call this characteristic temperature $T^{mid}$. Below the $T^{mid}$, the value of the $d\rho/dT$ begins to increase rapidly. At the same time, the resistivity begins to decrease largely, which indicates the evolution of the superconducting current path below the $T^{mid}$. Figure 6(b) displays the enlargement of temperature dependence of the resistivity under various magnetic fields up to 14 T. The $T^{mid}$ estimated from the $d\rho/dT$-$T$ curves using the method same as in Fig. 6(a) is indicated by arrows in Fig. 6(b). With increasing magnetic fields, the anomaly at $T^{mid}$ in the $\rho$-$T$ curve is more pronounced. Then, the $T^{mid}$ overlaps with the $T_c^{onset}$ under magnetic fields above 10 T. It can be considered that the $T_c^{onset}$, which denotes the evolution of superconducting current path, corresponds to the $T^{max}$ under lower magnetic fields while it corresponds to the $T^{mid}$ under higher magnetic fields. Hence, the anomalous behavior of the $\mu_0H_{c2}$-$T$ curve was observed in Fig. 4.

In Fig. 7(a), we plotted the $\mu_0H_{c2}^{max}(T)$, $\mu_0H_{c2}^{mid}(T)$ and $\mu_0H_{c2}^{min}(T)$ with the data points of the $\mu_0H_{c2}$ ($\rho$-$T$), $\mu_0H_{irr}$ ($\rho$-$T$) and $\mu_0H_{c2}$ (pulse). The $\mu_0H_{c2}^{min}(T)$ is obviously lower than the $\mu_0H_{c2}^{max}$ and $\mu_0H_{c2}^{mid}$ at whole temperatures, and it is located near $H_{irr}(T)$. As mentioned above, the low $\mu_0H_{c2}^{min}$ can be understood by the large anisotropy of superconductivity of LaO$_{0.5}$F$_{0.5}$BiS$_2$. Namely, the superconducting current path can be easily destroyed under magnetic fields for the grains with a grain orientation of $H // c$. Next, we discuss the difference between the $\mu_0H_{c2}^{max}$ and $\mu_0H_{c2}^{mid}$. As shown in Fig. 7(a), the curve of temperature dependence of the $\mu_0H_{c2}$ ($\rho$-$T$) almost overlaps with the $\mu_0H_{c2}^{max}(T)$ curve under magnetic fields up to 5 T. Then, it deviates from the $\mu_0H_{c2}^{max}(T)$ curve. Interestingly, the curve of the temperature dependence of the $\mu_0H_{c2}$ ($\rho$-$T$) finally overlaps with the $\mu_0H_{c2}^{mid}(T)$ curve under magnetic fields at above 10 T. The observed anomalous behavior of the $\mu_0H_{c2}$-$T$ curve implies that the superconducting current path is generated below $T^{max}$ under lower magnetic fields while it is generated below $T^{mid}$ under higher magnetic fields. Therefore, there should be a factor, which affects upper critical field, except for the difference in the grain orientation between $H//ab$ and $H//c$.

To explain the difference of two kinds of upper critical field, $\mu_0H_{c2}^{max}$ and $\mu_0H_{c2}^{mid}$, we assume that the superconducting states within the $ab$ plane are anisotropic. The structure distortion (or strain) could exist in the present sample of LaO$_{0.5}$F$_{0.5}$BiS$_2$ prepared using the high-pressure synthesis technique, as observed in the previous high-pressure studies on superconductivity and crystal structure of LaO$_{0.5}$F$_{0.5}$BiS$_2$ [23]. Although the evidence of lowering of symmetry of the crystal structure was not detected in the x-ray diffraction measurements, the broadening of the ($h$00) and (00$l$) peaks was observed in the high-pressure LaO$_{0.5}$F$_{0.5}$BiS$_2$ samples [25,26]. Therefore, it is conceivable that the crystal structure of the present LaO$_{0.5}$F$_{0.5}$BiS$_2$ polycrystalline sample is not a perfect tetragonal structure. If the $ab$ plane is distorted (or strained), the superconducting states along the $a$ axis could be non-equivalent to that along the $b$ axis. In that case, the upper critical field for the $a$-axis direction and the $b$-axis direction could be different due to the difference of effective mass. In addition, a previous theoretical study suggested that the one-dimensional nature of the Bi bands (Bi-6$p_x$ and Bi-6$p_y$) provides good nesting of the Fermi surface in the LaO$_{1-x}$F$_x$BiS$_2$ system [14]. Furthermore, Suzuki et al. showed that the band structure of the BiS$_2$ family could be largely affected by the change in the local crystal structures [15]. These theoretical studies could suggest a possibility of the appearance of the anisotropic superconducting states within the $ab$ plane when the degeneracy of Bi-6$p_x$ and Bi-6$p_y$ is lifted. Therefore, it is conceivable that there are the different upper critical fields of $\mu_0 H_{C2}^{a\text{-axis}}$ and $\mu_0 H_{C2}^{b\text{-axis}}$ in a distorted (or strained) Bi-S plane of LaO$_{0.5}$F$_{0.5}$BiS. On the basis of the discussion, the evolution of the anisotropic superconducting states of LaO$_{0.5}$F$_{0.5}$BiS$_2$ under magnetic fields is summarized as below. All the grains are superconducting at $\mu_0 H < \mu_0 H_{C2}^{\min}$. At magnetic fields of $\mu_0 H_{C2}^{\min} < \mu_0 H < \mu_0 H_{C2}^{\text{mid}}$, superconducting states of the grains with a grain orientation of $H \parallel c$ are suppressed. Then, under $\mu_0 H_{C2}^{\text{mid}} < \mu_0 H < \mu_0 H_{C2}^{\max}$, the superconducting states of the grains with an orientation of $H \parallel a$ or $H \parallel b$ are suppressed; we cannot distinguish the $a$-axis orientation from the $b$-axis orientation within the results of present study. Above $\mu_0 H_{C2}^{\max}$, all the grains become non-superconducting.

To discuss the anisotropy of upper critical field, we have plotted the values of the $\mu_0 H_{C2}^{\max}(T)$, $\mu_0 H_{C2}^{\text{mid}}(T)$ and $\mu_0 H_{C2}^{\min}(T)$ in Fig. 7(b). The data points for the $\mu_0 H_{C2}^{\text{mid}}$ and $\mu_0 H_{C2}$ (pulse) exhibit a linear temperature dependence. Therefore, we estimated the $d\mu_0 H_{C2}^{\max}/dT$, $d\mu_0 H_{C2}^{\text{mid}}/dT$ and $d\mu_0 H_{C2}^{\min}/dT$ by fitting the data points with a linear function of temperature to discuss the anisotropy. The estimated $d\mu_0 H_{C2}^{\max}/dT$, $d\mu_0 H_{C2}^{\text{mid}}/dT$ and $d\mu_0 H_{C2}^{\min}/dT$ are -7.05 (T/K), -2.60 (T/K) and -0.95 (T/K), respectively. The anisotropy parameter estimated using the values of the $d\mu_0 H_{C2}^{\max}/dT$ and $d\mu_0 H_{C2}^{\min}/dT$ is 7.4. The anisotropy parameter estimated in this work are clearly lower than the anisotropy parameter of over 30 for LaO$_{1-x}$F$_x$BiS$_2$ single crystals ($T_c \sim 3$ K) grown at ambient pressure [28]. This indicates that the larger anisotropy of superconductivity is not essential for the enhancement of the $T_c$, and the lower anisotropy may be important for a higher $T_c$ in LaO$_{1-x}$F$_x$BiS$_2$. Furthermore, the evolution of the in-plane anisotropy within Bi-S plane may play an important role for the enhancement of $T_c$. To elucidate the mechanism of superconductivity in the BiS$_2$ family, high-field studies using LaO$_{1-x}$F$_x$BiS$_2$ single crystals with higher $T_c$ are needed. To achieve it, single crystal growth of the high-$T_c$ phase of LaO$_{1-x}$F$_x$BiS$_2$ is required.

4. Conclusion

We have studied magnetic and transport properties of the high-pressure-synthesized LaO$_{0.5}$F$_{0.5}$BiS$_2$ poly crystalline sample under high magnetic fields. The temperature dependence of upper critical field shows anomalous behavior with a characteristic magnetic field of 8 T. We found that the anisotropy of upper critical field of the polycrystalline sample of LaO$_{0.5}$F$_{0.5}$BiS$_2$ can be estimated by analyzing the characteristic temperatures in the $d\rho/dT$-$T$ curve. It was found that there were three kinds of upper critical field in LaO$_{0.5}$F$_{0.5}$BiS$_2$. The lowest upper critical field $\mu_0 H_{C2}^{min}$ is regarded as the $\mu_0 H_{c2}$ when the magnetic fields are applied parallel to the $c$ axis of the grain. The $\mu_0 H_{C2}^{max}$ and $\mu_0 H_{C2}^{mid}$ are regarded as the $\mu_0 H_{c2}$ where the superconducting states are suppressed for the grains with an orientation of $H // ab$. The difference between the $\mu_0 H_{C2}^{max}$ and $\mu_0 H_{C2}^{mid}$ could be explained by the anisotropy of superconducting states within the Bi-S plane, which is possibly caused by the lattice distortion or strain within the Bi-S plane. Hence, the upper critical fields $\mu_0 H_{C2}^{a\text{-axis}}$ and $\mu_0 H_{C2}^{b\text{-axis}}$ are different due to the difference of effective mass. The fraction of the grains in superconducting states decreases with increasing magnetic field in the order of 0 T, $\mu_0 H_{C2}^{min}$, $\mu_0 H_{C2}^{mid}$ and $\mu_0 H_{C2}^{max}$. Above $\mu_0 H_{C2}^{max}$, all the grains become non-superconducting. The estimated $d\mu_0 H_{C2}^{max}/dT$, $d\mu_0 H_{C2}^{mid}/dT$ and $d\mu_0 H_{C2}^{min}/dT$ are -7.05 (T/K), -2.60 (T/K) and -0.95 (T/K), respectively. The anisotropy parameter for upper critical field estimated using $d\mu_0 H_{C2}^{max}/dT$ and $d\mu_0 H_{C2}^{min}/dT$ is 7.4, which is clearly lower than the anisotropy parameter of over 30 in the previous report for LaO$_{1-x}$F$_x$BiS$_2$ single crystals ($T_c \sim 3$ K) grown at ambient pressure. This indicates that the larger anisotropy of superconductivity is not essential for the enhancement of $T_c$, and the lower anisotropy might be important for a higher $T_c$ in LaO$_{1-x}$F$_x$BiS$_2$. Moreover, the evolution of the in-plane anisotropy within the Bi-S plane may play an important role for the enhancement of $T_c$ in LaO$_{1-x}$F$_x$BiS$_2$.

Acknowledgements

The authors would like to thank Prof. K. Kuroki of Osaka University for fruitful discussion. This work was partly supported by JSPS KAKENHI Grant Numbers 25707031, 23340096.

References

[1] J. G. Bednorz, and K. A. Müller, Z. Physik B 64, 189 (1986).

[2] Y. kamihara, T. Watanabe, M. Hirano, and H. Hosono, J. Am. Chem. Soc. 130, 3296 (2008).

[3] Y. Mizuguchi, H. Fujihisa, Y. Gotoh, K. Suzuki, H. Usui, K. Kuroki, S. Demura, Y. Takano, H. Izawa, and O. Miura, Phys. Rev. B. 86, 220510 (2012).

[4] Y. Mizuguchi, S. Demura, K. Deguchi, Y. Takano, H. Fujihisa, Y. Gotoh, H. Izawa, and O. Miura, J. Phys. Soc. Jpn. 81, 114725 (2012).

[5] J. Xing, S. Li, X. Ding, H. Yang, and H. H. Wen, Phys. Rev. B. 86, 214518 (2012).

[6] S. Demura, K. Deguchi, Y. Mizuguchi, K. Sato, R. Honjyo, A. Yamashita, T. Yamaki, H. Hara, T. Watanabe, S. J. Denholme, M. Fujioka, H. Okazaki, T. Ozaki, O. Miura, T. Yamaguchi, H. Takeya, and Y. Takano, arXiv:1311.4267.

[7] R. Jha, A. Kumar, S. K. Singh, and V. P. S. Awana, J. Sup. Novel Mag. 26, 499 (2013).

[8] J. Kajitani, K. Deguchi, T. Hiroi, A. Omachi, S. Demura, Y. Takano, O. Miura, and Y. Mizuguchi, J. Phys. Soc. Jpn., in printing (arXiv: 1401.7506).

[9] S. Demura, Y. Mizuguchi, K. Deguchi, H. Okazaki, H. Hara, T. Watanabe, S. J. Denholme, M. Fujioka, T. Ozaki, H. Fujihisa, Y. Gotoh, O. Miura, T. Yamaguchi, H. Takeya, and Y. Takano, J. Phys. Soc. Jpn. 82, 033708 (2013).

[10] R. Jha, A. Kumar, S. K. Singh, and V. P. S. Awana, J. Appl. Phys. 113, 056102 (2013).

[11] D. Yazici, K. Huang, B. D. White, A. H. Chang, A. J. Friedman, and M. B. Maple, Philosophical Magazine 93, 673 (2012).

[12] X. Lin, X. Ni, B. Chen, X. Xu, X. Yang, J. Dai, Y. Li, X. Yang, Y. Luo, Q. Tao, G. Cao, and Z. Xu, Phys. Rev. B 87, 020504 (2013).

[13] H. Sakai, D. Kotajima, K. Saito, H. Wadati, Y. Wakisaka, M. Mizumaki, K. Nitta, Y. Tokura, and S. Ishiwata, J. Phys. Soc. Jpn. 83, 014709 (2014).

[14] H. Usui, K. Suzuki, and K. Kuroki, Phys. Rev. B 86, 220501 (2012).

[15] K. Suzuki, H. Usui, and K. Kuroki, Phys. Procedia, 45, 21 (2013).

[16] T. Yildirim, Phys. Rev. B 87, 020506(R) (2013).

[17] I. R. Shein and, A. L. Ivanovskii, JETP Lett. 96, 859 (2012).

[18] B. Li, Z. W. Xing, and G. Q. Huang, arXiv:1210.1743.

[19] G. Martins, A. Moreo, and E. Dagotto, Phys. Rev. B 87, 081102 (2013).

[20] C. Morice, E. Artacho, S. E. Dutton, D. Molnar, H. J. Kim, and S. S. Saxena, arXiv:1312.2615.

[21] H. Kotegawa, Y. Tomita, H. Tou, H. Izawa, Y. Mizuguchi, O. Miura, S. Demura, K. Deguchi, and Y. Takano, J. Phys. Soc. Jpn, 81, 103702 (2012).
8

Figure captions

Fig. 1. (a) Temperature dependence of magnetic susceptibility for the polycrystalline sample of LaO$_{0.5}$F$_{0.5}$BiS$_2$ prepared using the high-pressure synthesis method. (b) Temperature dependence of FC susceptibility for LaO$_{0.5}$F$_{0.5}$BiS$_2$ with applied fields of 0.5 mT, 0.5 T and 1 T.

Fig. 2. (a) Temperature dependence of electrical resistivity for LaO$_{0.5}$F$_{0.5}$BiS$_2$ under magnetic fields up to 14 T. (b) The enlargement of the temperature dependence of electrical resistivity for LaO$_{0.5}$F$_{0.5}$BiS$_2$ under magnetic fields up to 14 T. The dashed line indicates $T_c^{onset}$.

Fig. 3. Magnetic field dependence of electrical resistivity for LaO$_{0.5}$F$_{0.5}$BiS$_2$ at temperatures of 1.4 K, 2.4 K, 3.4 K and 4.2 K.

Fig. 4. Temperature dependence of $\mu_0 H_{C2}$ ($\rho$-$T$), $\mu_0 H_{irr}$ ($\rho$-$T$) and $\mu_0 H_{C2}$ (pulse).

Fig. 5. (a-d) Temperature dependence of $dM/dT$ and $d\rho/dT$ for LaO$_{0.5}$F$_{0.5}$BiS$_2$ under applied magnetic fields of (a) 0.5 T, (b) 1 T, (c) 2 T and (d) 3 T. In the panel (b), the $T^{max}$ and $T^{min}$ are indicated with arrows.

Fig. 6. (a) Temperature dependence of resistivity ($\rho$) and $d\rho/dT$ LaO$_{0.5}$F$_{0.5}$BiS$_2$ under a magnetic field of 8 T. The $T^{max}$ and $T^{mid}$ are indicated with dashed lines and arrows. The $T_c^{onset}$ is the temperature where resistivity begins to decrease. (b) The enlargement of the temperature dependence of electrical resistivity for LaO$_{0.5}$F$_{0.5}$BiS$_2$ under magnetic fields up to 14 T. The $T^{mid}$ is indicated with arrows.

Fig. 7 (a) Magnetic field-Temperature phase diagram of LaO$_{0.5}$F$_{0.5}$BiS$_2$ established with the estimated $\mu_0 H_{C2}$ (pulse), $\mu_0 H_{C2}^{max}$, $\mu_0 H_{C2}^{mid}$ and $\mu_0 H_{C2}^{min}$. The $\mu_0 H_{C2}$ ($\rho$-$T$) and $\mu_0 H_{irr}$ ($\rho$-$T$) are also plotted. In the colored area, below $T_c^{onset}$ ($\rho$-$T$), the path of the superconducting current is generated. (b) Magnetic field-Temperature phase diagram of LaO$_{0.5}$F$_{0.5}$BiS$_2$ with three kinds of upper critical fields, $\mu_0 H_{C2}^{max}$, $\mu_0 H_{C2}^{mid}$ and $\mu_0 H_{C2}^{min}$. The dashed lines are the linear fitting lines for $\mu_0 H_{C2}^{max}$, $\mu_0 H_{C2}^{mid}$ and $\mu_0 H_{C2}^{min}$.

Fig. 1.

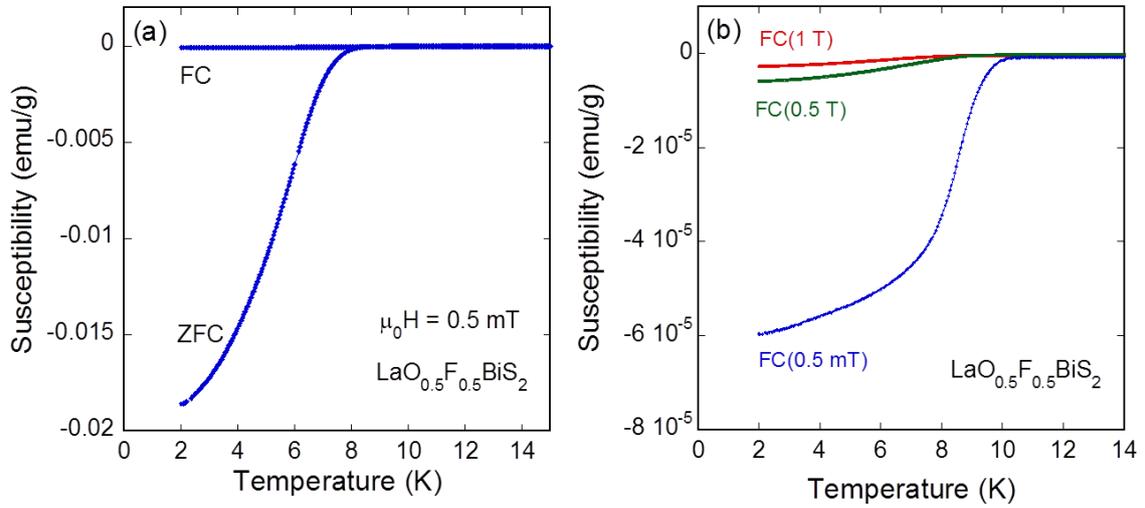

Fig. 2.

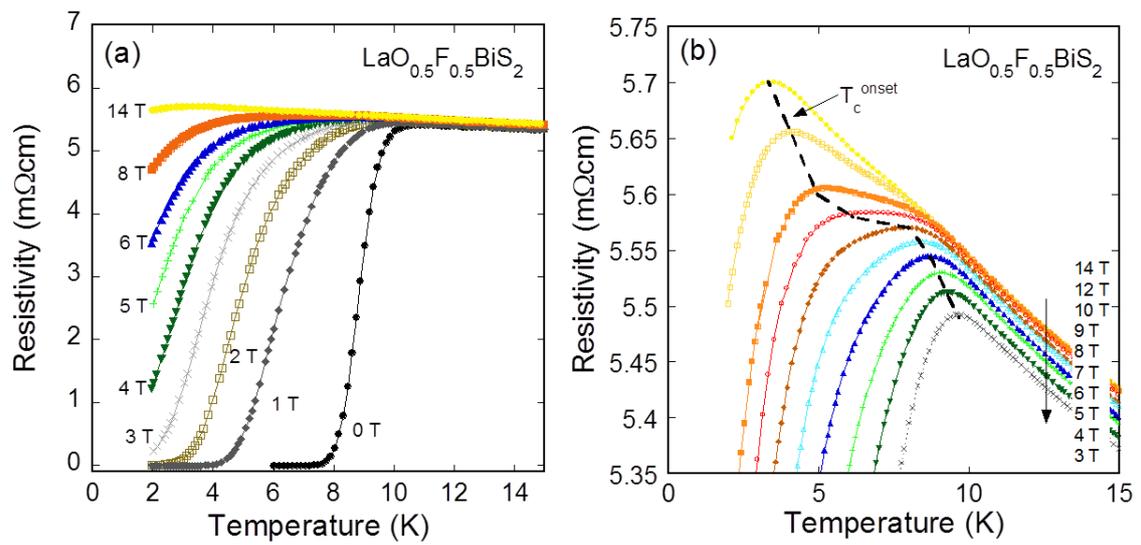

Fig. 3.

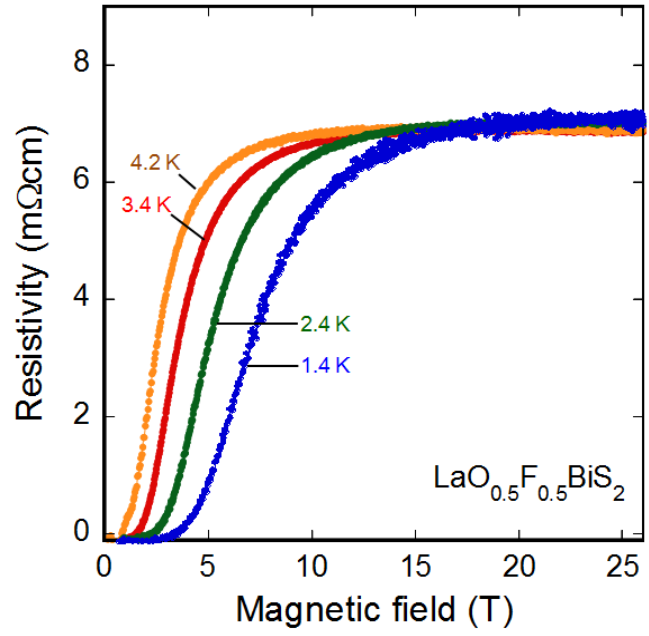

Fig. 4.

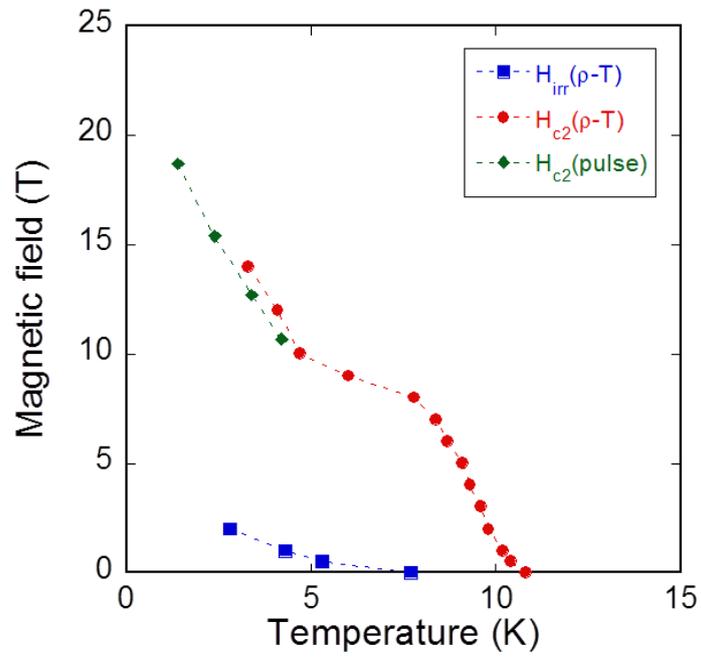

Fig. 5.

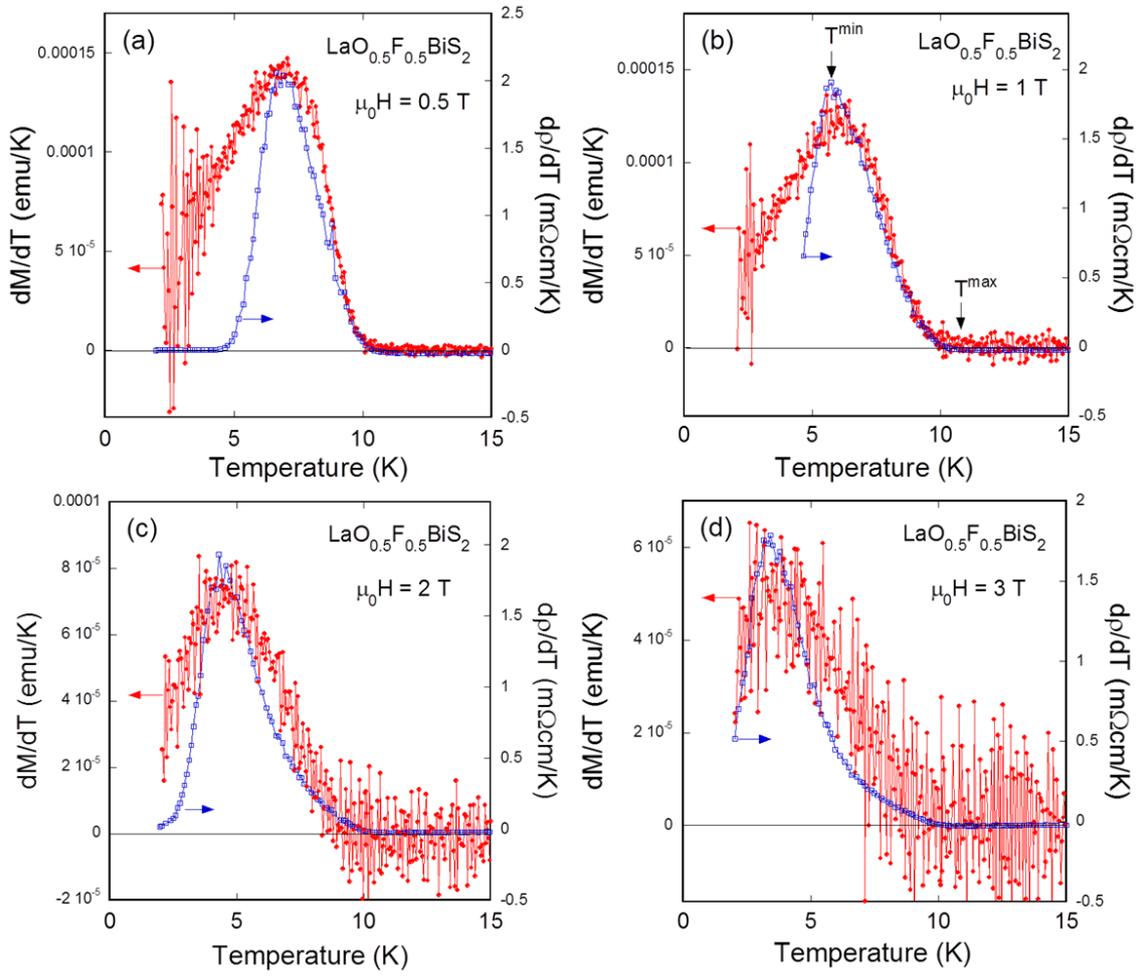

Fig. 6.

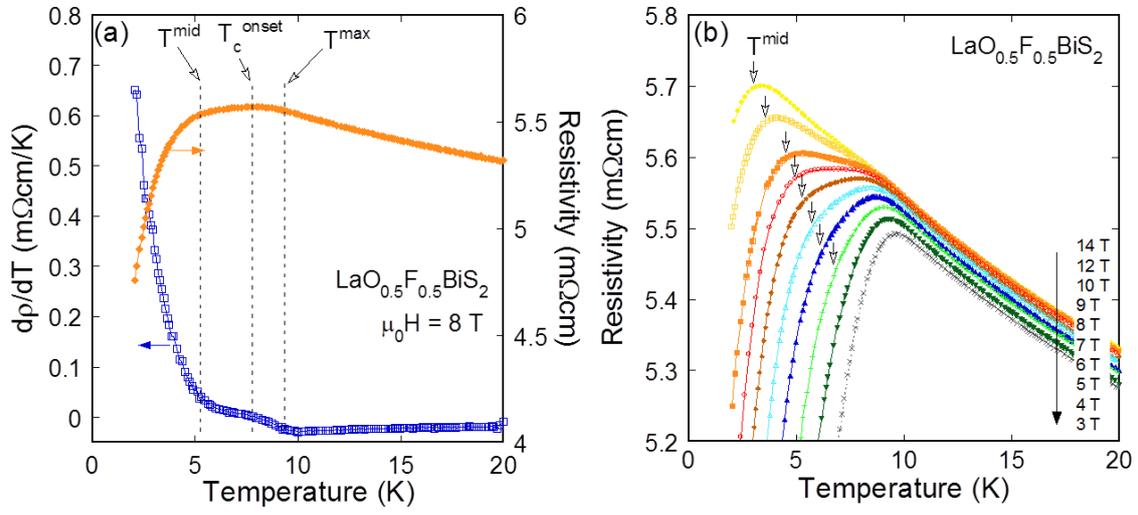

Fig. 7.

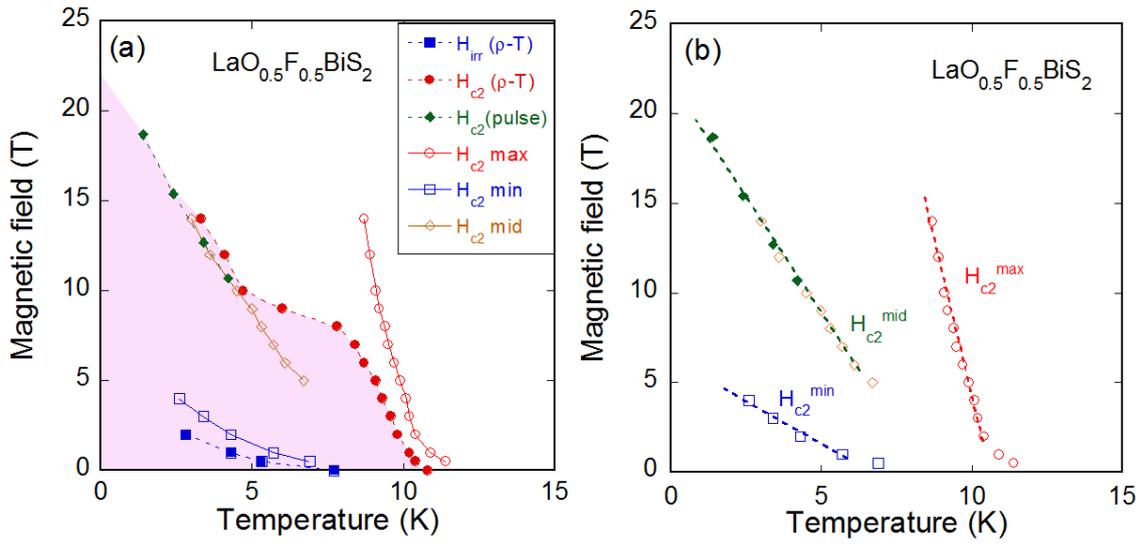